Table I. Interatomic distances along the chain of several PtCl complexes, general formula $[Pt(L-L)_2][Pt(L-L)_2Cl_2]Y_4$ (from Refs. 2-4).

|     | L-L[a]    | Y       | $d(Pt^{IV}-Cl)$/Å | $d(Pt^{II}-Cl)$/Å | $d(Pt^{IV}-Pt^{IV})$/Å | $r$[b] |
|-----|-----------|---------|-------------------|-------------------|------------------------|--------|
| (1) | chxn      | $ClO_4$ | 2.315             | 3.396             | 5.711                  | 0.68   |
|     | pn        | $ClO_4$ | 2.29              | 3.22              | 5.512                  | 0.71   |
|     | $(NH_3)_2$| $HSO_4$ | 2.310             | 3.158             | 5.466                  | 0.73   |
| (2) | en        | $ClO_4$ | 2.327             | 3.101             | 5.442                  | 0.75   |
|     | etn       | Cl      | 2.26              | 3.13              | 5.39                   | 0.72   |
|     | en        | $CuCl_4$| 2.328             | 2.937             | 5.261                  | 0.79   |
| (3) | chxn      | Cl      | 2.324             | 2.834             | 5.158                  | 0.82   |

[a]Ligand: chxn = (−)(R,R)-1,2-diaminecyclohexane; pn = 1,2-diaminepropane; en = 1,2-diamine-ethane (ethylendiamine); etn = ethylamine.

[b]Ratio between the $Pt^{IV}$-Cl and $Pt^{II}$-Cl distances.

Table II. Force constants for the isolated model compound.

| Symbol[a] | Description of coordinates involved | Value[b] |
|---|---|---|
| $K_1$ | Pt-N stretch | 2.60 |
| $K_2$ | N-C stretch | 5.70 |
| $K_3$ | Pt-Cl stretch | 2.00 |
| $H_1$ | $N_a PtN_b$ bend | 3.80 |
| $H_2$ | $N_a PtN_c$ bend | 0.84 |
| $H_3$ | NPtCl bend | 1.05 |
| $F_1$ | Pt-$N_a$ stretch / Pt-$N_d$ stretch | 0.63 |
| $F_2$ | Pt-$N_a$ stretch / Pt-$N_b$ stretch | 0.35 |
| $F_3$ | Pt-$N_a$ stretch / Pt-$N_c$ stretch | 0.10 |
| $F_4$ | Pt-N stretch / N-C stretch | 0.50 |
| $F_5$ | Pt-$Cl_a$ stretch / Pt-$Cl_b$ stretch | 0.25 |

[a] K, stretching force constants; H, bending force constants; F, off-diagonal (interaction) force constants.
[b] K and F force constants expressed in mdyn/A; H force constants expressed in mdyn A/(rad)$^2$.

Table III. Stretching force constants (mdyn/Å) along the Metal-Halogen chain for the three studied compounds.

| Symbol | Coordinates involved | (1) | (2) | (3) |
|---|---|---|---|---|
| $K_3$ | $Pt^{IV}$-Cl stretch | 2.00 | 2.00 | 2.00 |
| $K_4$ | $Pt^{II}$-Cl stretch | 0.33 | 0.40 | 0.55 |
| $F_5$ | $Pt^{IV}$-Cl stretch / $Pt^{IV}$-Cl stretch | 0.25 | 0.25 | 0.25 |
| $F_6$ | $Pt^{IV}$-Cl stretch / $Pt^{II}$-Cl stretch | 0.12 | 0.12 | 0.12 |
| $F_7$ | $Pt^{II}$-Cl stretch / $Pt^{II}$-Cl stretch | 0.00 | 0.00 | 0.00 |

Table V. Assignment of the PtCl infrared spectra (500-100 cm$^{-1}$).

| Experimental frequency[a] | | | Calc. freq.[b] | Approx. description[c] |
|---|---|---|---|---|
| (1) | (2) | (3) | | |
| 439 m | 473 w | 439 m | 456–457 | $\nu$(Pt–N) |
| 364 sh | 357 sh | 385 w | 356–357 | $\delta$(NPtN) |
|  |  | 371 m |  |  |
| 352 s | 355 s | 362 m | See Table V | $\nu$(Pt–Cl) ($\omega_a$) |
|  | 352 sh |  |  |  |
| 287 m | 292 s | 293 m | 269–272 | $\delta$(ClPtN) |
| 260 sh | 289 sh |  |  |  |
| 254 m | 253 s | 272 m | 249–252 | $\delta$(ClPtN) |
|  | 238 sh |  |  |  |
| 184 w | 188 w,br | 182 w | 183–194 | $\delta$(ClPtN) |
| 175 w |  |  |  |  |
| 154 w | 139 s | 127 s | --- | "external" ligand? |
| 100 w |  | 111 s | --- |  |

[a] Frequencies in cm$^{-1}$. Approximate relative intensities given by: s, strong; m, medium; w, weak; sh, shoulder; br, broad.

[b] Frequency interval (cm$^{-1}$) in the calculations for the three compounds.

[c] The conventional spectroscopic notation is used: $\nu$ stands for stretching mode, $\delta$ for bending.

Table IV. Reference and experimental longitudinal frequencies (cm$^{-1}$).

| Complex | $\omega_s^\circ$ | $\omega_s$(obs.) | $\omega_a^\circ$ | $\omega_a$(obs.) | $\omega_3^\circ$ |
|---|---|---|---|---|---|
| **(1)** | 335 | 327 | 352 | 352 | 85 |
| **(2)** | 340 | 312 | 355 | 355 | 92 |
| **(3)** | 350 | 290 | 362 | 362 | 104 |

Table VI. Microscopic parameters and optical data of the three studied compounds (experimental data from present work and Refs. 2-4).

---

$[Pt(chxn)_2][Pt(chxn)_2Cl_2](ClO_4)_4$ **(1)**

| | | | |
|---|---|---|---|
| $g_S$ = 0.172 eV | $\omega_S^o$ = 335 cm$^{-1}$ | $t$ = 0.27 eV | $\sigma$ = 0.96 |

$\omega_S$ (calc) = 327 cm$^{-1}$ ; $\omega_S$ (expt) = 327 cm$^{-1}$
$\chi_V$ (calc) = 0.066 eV$^{-1}$ ; $\chi_V$ (expt) = 0.066 eV$^{-1}$
$\omega_{CT}$(calc) = 3.1 eV ; $\omega_{CT}$(expt) = 3.2 eV
$f_{CT}$(calc) = 1.2 ; $f_{CT}$(expt) = ---
$\omega_L$ (calc) = 1.86 eV ; $\omega_L$ (expt) = 1.49 eV at 2 K
$[I(2\omega_S)/I(\omega_S)]$ (calc) = 0.56 ; $[I(2\omega_S)/I(\omega_S)]$ (expt) = ---

---

$[Pt(en)_2][Pt(en)_2Cl_2](ClO_4)_4$ **(2)**

| | | | |
|---|---|---|---|
| $g_S$ = 0.188 eV | $\omega_S^o$ = 340 cm$^{-1}$ | $t$ = 0.40 eV | $\sigma$ = 0.87 |

$\omega_S$ (calc) = 312 cm$^{-1}$ ; $\omega_S$ (expt) = 312 cm$^{-1}$
$\chi_V$ (calc) = 0.188 eV$^{-1}$ ; $\chi_V$ (expt) = 0.188 eV$^{-1}$
$\omega_{CT}$(calc) = 2.68 eV ; $\omega_{CT}$(expt) = 2.72 eV
$f_{CT}$(calc) = 3.1 ; $f_{CT}$(expt) = 3.0
$\omega_L$ (calc) = 1.56 eV ; $\omega_L$ (expt) = 1.22 eV at 2 K
$[I(2\omega_S)/I(\omega_S)]$ (calc) = 0.45 ; $[I(2\omega_S)/I(\omega_S)]$ (expt) = 0.61, $\lambda_o$=2.41eV, 80K

---

$[Pt(chxn)_2][Pt(chxn)_2Cl_2]Cl_4$ **(3)**

| | | | |
|---|---|---|---|
| $g_S$ = 0.206 eV | $\omega_S^o$ = 350 cm$^{-1}$ | $t$ = 0.44 eV | $\sigma$ = 0.78 |

$\omega_S$ (calc) = 290 cm$^{-1}$ ; $\omega_S$ (expt) = 290 cm$^{-1}$
$\chi_V$ (calc) = 0.321 eV$^{-1}$ ; $\chi_V$ (expt) = 0.321 eV$^{-1}$
$\omega_{CT}$(calc) = 2.24 eV ; $\omega_{CT}$(expt) = 2.19 eV
$f_{CT}$(calc) = 4.5 ; $f_{CT}$(expt) = ---
$\omega_L$ (calc) = 1.26 eV ; $\omega_L$ (expt) = 0.91 eV at 2 K
$[I(2\omega_S)/I(\omega_S)]$ (calc) = 0.45 ; $[I(2\omega_S)/I(\omega_S)]$ (expt) = 0.64, $\lambda_o$=2.71eV, 80K

FIGURES

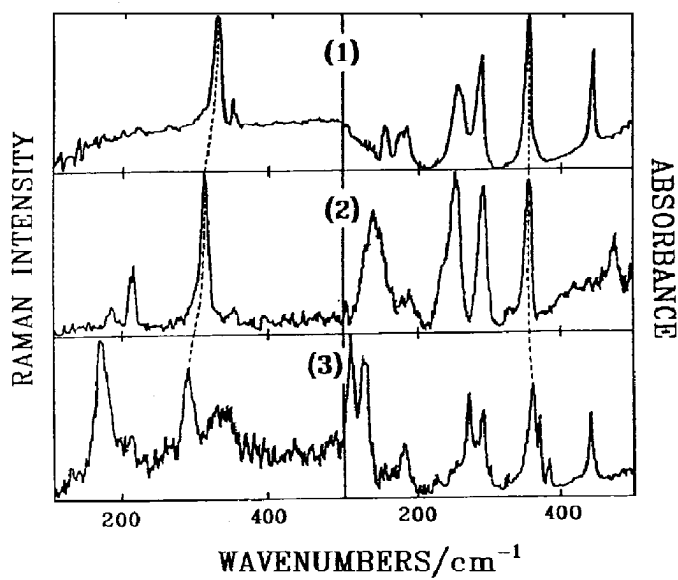

Fig. 1 – Raman (left panels) and infrared (right panels) spectra of: [Pt(chxn)$_2$][Pt(chxn)$_2$Cl$_2$](ClO$_4$)$_4$, **(1)**; [Pt(en)$_2$][Pt(en)$_2$Cl$_2$](ClO$_4$)$_4$, **(2)**; and [Pt(chxn)$_2$][Pt(chxn)$_2$Cl$_2$]Cl$_4$, **(3)**, from 100 to 500 cm$^{-1}$. The dashed lines put in evidence the different behavior of the symmetric (Raman) and antisymmetric (infrared) Pt-Cl stretching with decreasing Pt-Pt distance from **(1)** to **(3)**.

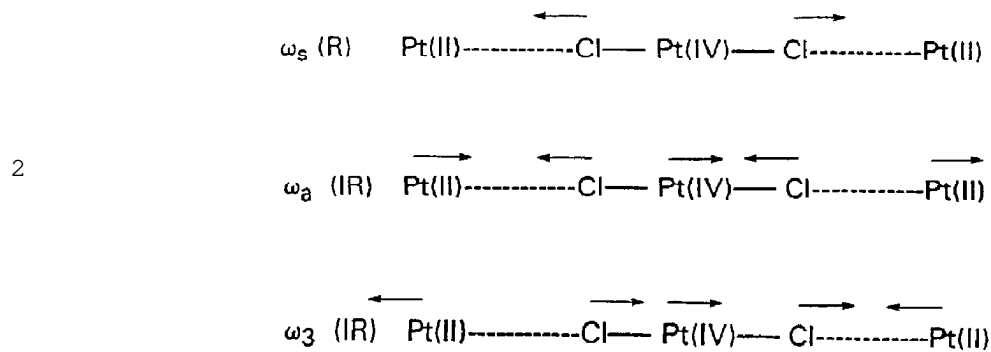

Fig. 2 – Approximate description of the zone center PtCl longitudinal normal modes (from Ref. 4).

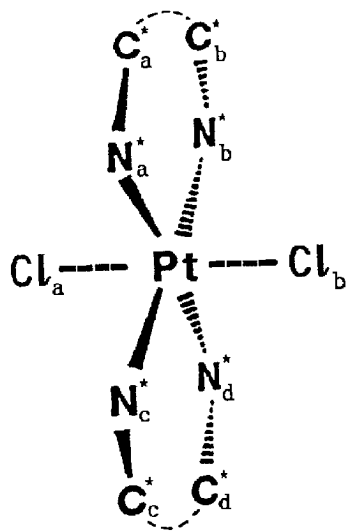

Fig. 3 - Schematic representation of the model Pt$^{IV}$ complex used in the normal coordinate analysis. The same structure, without the Cl atoms, is adopted for the Pt$^{II}$ complex.

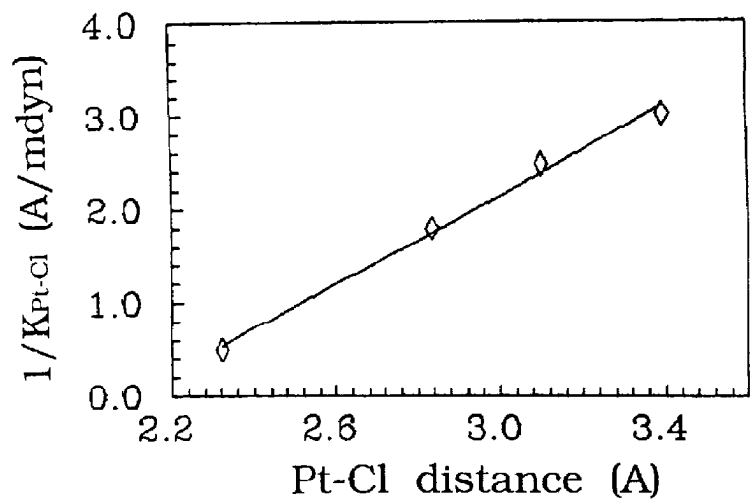

Fig. 4 - Plot of the reciprocal Pt-Cl stretching force constants used in the normal coordinate analysis (Table III, $K_3$ and $K_4$) vs. the Pt-Cl distance.

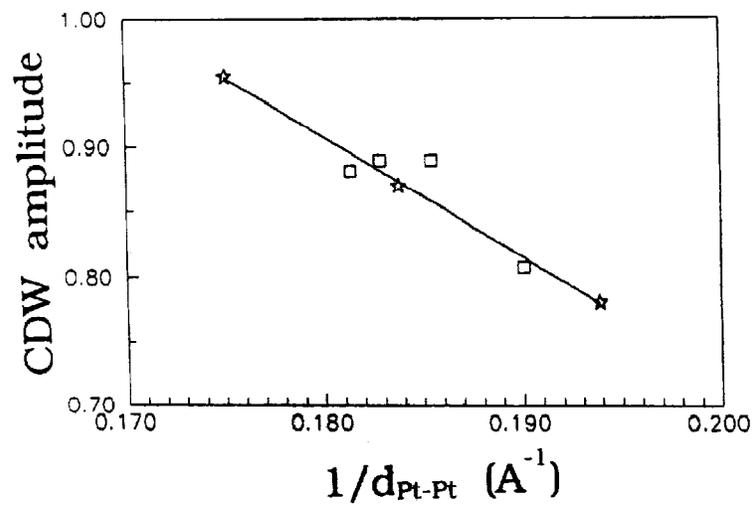

Fig. 5 – CDW amplitude estimated for the three studied compounds (stars) vs. the reciprocal Pt-Pt distance. The squares refer to the other PtCl complexes listed in Table I.

# MIXED-VALENCE HALOGEN-BRIDGED Pt COMPLEXES

Alberto Girlando, Anna Painelli, and Marco Ardoino*
Dip. C.G.I.C.A. e Chimica Fisica, Parma University
viale delle Scienze, 43100 Parma, Italy

Carlo Bellitto
Ist. Chimica dei Materiali, C.N.R.
Monterotondo Staz. 10, Roma, Italy

The spectroscopic effects of electron-phonon (e-ph) coupling in mixed- valence halogen-bridged chain complexes are investigated through a parallel infrared and Raman study of three compounds with decreasing Metal-Metal distance along the chain. The e-ph interaction is analyzed in terms of the Herzberg-Teller coupling scheme. We take into account the quadratic term and define a precise reference state. The force field relevant to this state is constructed, whereas the electronic structure is analyzed in terms of a simple phenomenological model, singling out a trimeric unit along the chain. In this way we are able to account for all the available optical data of the three compounds, and to estimate the relevant microscopic parameters, such as the e-ph coupling constants and the Charge Density Wave (CDW) amplitude. The latter correlates linearly with the reciprocal Metal- Metal distance.

PACS: $71.45.Lr; 63.20.Kr; 78.30.-j; 71.38.+i$

*Present address: Istituto M.A.S.P.E.C.- C.N.R., via Chiavari 18/a, 43100 Parma, Italy.

## I. INTRODUCTION

Halogen-bridged linear-chain transition-metal complexes (MX chains) are characterized by either a Charge Density Wave (CDW) or a Spin Density Wave (SDW) ground state, the former being the most common.[1-3] In the chemical literature[4] MX chains are indeed referred to as "mixed valence" complexes, a typical formula being: $[M^{(IV-\rho)}(L-L)_2X_2][M^{(II+\rho)}(L-L)_2]Y_4$, with $M = Pt, Pd, Ni$; X: an halogen ion; $L-L$: a bidentate amine ligand; Y: a closed-shell negative counterion, $ClO_4^-, Cl^-$, etc. The amplitude of the CDW is $\sigma = 1 - \rho$, and it can be tuned[2,3] by changing the bridging halide, the ligand and/or the counter-ion, thus inducing changes in the crystal packing and/or in the hydrogen bond strength. This tunability, together with their crystalline nature, makes MX chains useful model systems for testing theories of one-dimensional (1D) solids and in particular to study the interplay between electron-electron and electron-phonon (e-ph) couplings.[5,7] The interest in these compounds is further increased due to the observation of clear spectral signatures of defect states (polarons, solitons, etc.).[7]

On the other hand, whereas the role of e-ph coupling in determining the CDW ground state of MX chains has been perhaps overemphasized, its spectroscopic effects have not been fully appreciated. These effects are analogous to those observed in organic charge-transfer (CT) crystals and conjugated polymers,[8] formed on $[Pt(en)_2Cl_2][Pt(en)_2](ClO_4)_4$ (en = ethylendiammine) under pressure have put in evidence some spectroscopic consequences of e-ph coupling. In the present paper we use the "chemical" rather than the physical pressure to study the spectroscopic effects following a controlled change in the CDW strength. In other words, we analyze the ambient pressure Raman and IR data relevant to three members of the so-called PtCl series:

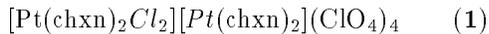  (1)

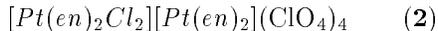  (2)

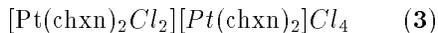  (3)

(chxn: $(-)(R,R)-1,2$-diaminociclohexane). As shown in Table I, where the structural data of several PtCl complexes are reported, the ratio $r$ between $Pt^{IV} - Cl$ and $Pt^{II} - Cl$ distances smoothly increases from (1) to (3), in correspondence with a decrease of the $Pt^{IV} - Pt^{II}$ distance. By exploiting the known crystal data,[2,3] we construct a reference force field that, together with a careful analysis of e-ph coupling, allows us to rationalize the main vibrational features of the three compounds and to account for the apparently anomalous behavior of the Pt-Cl stretching frequencies under chemical or physical pressure. Furthermore, we show that Raman, absorption and luminescence spectra of the three compounds are rationalized in terms of a phenomenological model of the PtCl electronic structure, properly accounting for e-ph coupling. The fit to experimental data yields reliable estimate of the CDW amplitude. By relating the CDW amplitude of the three compounds with the corresponding structural data, we offer simple empirical criteria to evaluate the CDW amplitude from a limited set of experimental data for complexes within the PtCl series.

## II. EXPERIMENT AND RESULTS

The three PtCl compounds have been prepared as reported in the literature.[11,12] The IR spectra have been obtained by a Bruker IFS66 FT spectrometer, with Mylar beamsplitter and DTGS detector. For the Raman spectra we have used the same spectrometer, coupled with a Bruker photoluminescence module, using a quartz beamsplitter and Si detector. The excitation light has been the the 647 nm line of a LEXEL K1000 Kr ion laser. Rejection of the laser line has been achieved by a Kaiser Supernotch filter. The spectral resolution for both IR and Raman is 2 $cm^{-1}$.

Fig. 1 reports the Raman and IR spectra of the powders of the three compounds from 100 to 500 $cm^{-1}$. In the Figure we have put in evidence the bands due to the symmetric (Raman) and antisymmetric (IR) Pt-Cl stretching modes, on which most of the following discussion will be based. From the data in the Figure it is clear that by increasing the chemical pressure (i.e., by contracting the chain) the Pt-Cl symmetric ($\omega_s$) and antisymmetric ($\omega_a$) stretching frequencies have opposite behavior: The former decreases and the latter increases. The trend is the same as that observed by increasing the physical pressure.[10] As it will become evident in the following, this behavior is a consequence of the e-ph coupling.

## III. ELECTRON-PHONON COUPLING

The spectroscopic signatures of e-ph coupling in the vibrational spectra of 1D solids, notably CT organic crystals and conjugated polymers, have been analyzed by a number of authors.[13] We have shown[14] that be included to properly define the reference state and the linear e-ph coupling constants. To such aim, the analysis is conveniently carried out in terms of the Herzberg-Teller (HT) coupling scheme, that we will sketch below before specializing the approach to MX chains.

In the crude adiabatic approximation the electronic hamiltonian ($\mathcal{H}_e$) refers to the equilibrium nuclear configuration and does not depend on the nuclear coordinates (Q). The Q-dependence is introduced by expanding $\mathcal{H}_e$ on the Q basis. Consistent with the harmonic approximation, the expansion is carried out up to the second order. The e-ph hamiltonian reads:

$$\mathcal{H}_{e-ph} = \sum_m \frac{\partial \mathcal{H}_e}{\partial Q_m} Q_m + \frac{1}{2} \sum \frac{\partial^2 \mathcal{H}_e}{\partial Q_m \partial Q_n} Q_m Q_n \tag{1}$$

where m,n count the nuclear coordinates. $\mathcal{H}_{e-ph}$ acts as a perturbation on the Q-independent eigenstates of $\mathcal{H}_e$, yielding Q-dependent states. The second derivatives of the ground state energy (the HT nuclear potential) on the Q's give the force constant matrix ($\Phi$) for the nuclear motion:

$$\Phi_{mn} = \Phi_{mn}^s + < G \mid \frac{\partial^2 \mathcal{H}_e}{\partial Q_m \partial Q_n} \mid G > -2 \sum_F \frac{< G \mid \partial \mathcal{H}_e / \partial Q_m \mid F >< F \mid \partial \mathcal{H}_e / \partial Q_n \mid G >}{E_F - E_G} \tag{2}$$

where |G> and |F> are the ground and excited states of $\mathcal{H}_e$, with energies $E_G$ and $E_F$, respectively and $\Phi_{mn}^s$ represents the skeleton force constants due to the contribution of the core electrons, i.e. of electrons not described by $\mathcal{H}_e$.

The distinctive vibrational properties of 1D systems originate from the third term of Eq. 2, that is from the linear e-ph coupling, which involves CT (intervalence) electron fluctuations and is strongly dependent on the excitation spectrum. On the other hand, the contribution from quadratic coupling only involves the ground state and is therefore expected to be similar for systems with similar ground states. The quadratic contribution is conveniently included in a reference force field $\Phi_{mn}^o$, which describes the potential due to the core electrons and to the ground state CT-electron distribution, but does not include CT-electron fluctuations.[8]

To apply the above scheme to MX chains the relevant $\mathcal{H}_e$ has to be defined. Although the two-band model proposed by Tinka-Gammel et al.[15] has proved useful to evidentiate a number of subtle effects, we believe it is unnecessarily complex to account for the phenomena we are dealing with in this paper. So we adopt a one-band model, neglecting the mixing of Pt and Cl orbitals. In the spirit of the crystal field approximation, we assume that the halogen atoms, bearing a fixed $- \mid e \mid$ charge, only act as an electrostatic potential modifying the on-site Pt energies.[6,9] Considering just one $d$ orbital per Pt site, the electronic hamiltonian is written as:

$$\mathcal{H}_e = \sum_i \epsilon_i n_i + t \sum_{i,\sigma} (a_{i,\sigma}^+ a_{i+1,\sigma} + \text{ H.c.})$$

$$+ U \sum_i a_{i,\alpha}^+ a_{i,\beta}^+ a_{i,\beta} a_{i,\alpha} + \frac{1}{2} \sum_{i,j} {'} V_{ij} q_i q_j \tag{3}$$

operator for a spin $\sigma$ electron in the $i$-th site, $n_i = \sum_\sigma a^\dagger_{i,\sigma} a_{i,\sigma}$ is the number operator for the $i$-th site, and $q_i = 4 - n_i$ is the corresponding charge operator. Moreover, $\epsilon_i$ represents the energy on the $i$-th Pt site, $t$ is the CT integral between adjacent sites, and $U$ and $V_{ij}$ are the repulsion energies for two electrons residing on the same site or on sites $i$ and $j$.

We now turn attention to e-ph coupling. We neglect the coupling induced by the ligand vibrations. This coupling is presumably small, unless the ligand is a charged atom, as in the so-called neutral chains (e.g. $[Pt(en)_2Cl_2][Pt(en)_2Cl_4]$).[4] We are therefore left with the longitudinal modes of the chain. There are three optically active longitudinal modes: $Q_s$, approximately described as the symmetric $Pt^{IV} - Cl$ stretching, Raman active; $Q_a$, the corresponding antisymmetric mode, IR active; and $Q_3$, approximately described as a $Pt - Pt$ stretching, IR active. The form of the modes[4] is sketched in Fig. 2. Out of the two IR modes, $Q_a$ is decoupled from the electron system, whereas $Q_3$ modulates on-bond integrals like $t$ (on-bond e-ph coupling, $e - bph$). Due to the lack of reliable spectral data in the $\omega_3$ frequency region (see below), we do not discuss the corresponding coupling. On the other hand, due to symmetry reasons,[13] $Q_3$ does not interact with the Raman $Q_s$ mode. So in the following we will concentrate on the $Q_s$ mode, which is coupled to the electron via a modulation of the on-site energies (on-site e-ph coupling, $e - sph$). By including the effects of quadratic e-ph coupling in the reference state, the linear site-diagonal e-ph hamiltonian reads:

$$\mathcal{H}_{e-sph} = \sqrt{\frac{\omega_s^o}{N}} Q_s g_s \sum_i (-1)^i n_i \qquad (4)$$

where N is the total number of Metal sites. In the adopted crystal-field approach the modulation of on-site energies by the Pt-Cl stretchings is due to the modulation of the distance between the negative charges on Cl ions and the Pt site. The $e - sph$ coupling constant, $g_s$, can be calculated if the electrostatic potential is given. We consider an unscreened Coulomb potential, so that:[9]

$$g_s = \sqrt{\frac{h}{\omega_s^o m}} \frac{2e^2}{\pi \epsilon_o d^2} \frac{d}{d\xi} \left\{ \sum_{i \text{ odd}} (-1)^{(i+1)/2} \frac{\xi}{i^2 - \xi^2} \right\} \qquad (5)$$

where m is the reduced mass of $Q_s$ (equal to the Cl mass), $d$ the Pt-Pt distance, $e$ the electron charge, $\epsilon_o$ the dielectric constant, and $\xi = 2(mN)^{-1/2} Q_s d^{-1}$ measures the chain distortion.

With the $e - sph$ hamiltonian of Eq. 4, the frequency of the $e - sph$ coupled mode becomes (see Eq.2):

$$\omega_s^2 = (\omega_s^o)^2 (1 - \frac{\chi_v g_s^2}{\omega_s^o}) \qquad (6)$$

where $\omega_s^o$ is the frequency relevant to the reference state, and $\chi_v$ the electronic response to $e - sph$ perturbation:

$$\chi_v = \frac{2}{N} \sum_F \frac{|<G| \sum_i (-1)^i n_i | F>|^2}{E_F - E_G} \qquad (7)$$

We notice that $\chi_v$ is a purely electronic quantity, which can be experimentally determined from the Raman frequency of symmetric Pt-Cl stretching mode, provided that the corresponding reference frequency, $\omega_s^o$, is known. From the discussion following Eq. 2, it turns out that the reference state corresponds to a hypothetical chain with the same CDW amplitude as the actual chain, but with negligible fluctuations of procedure, illustrated in the next Section, is in many respects analogous to that we have adopted to construct a reference force field for trans- polyacetylene.[16]

## IV. REFERENCE FORCE FIELD

Several force fields have been proposed to account for the longitudinal vibrations of MX chains.[17–19] The problem looks rather simple, but many alternative choices are possible. A first systematic investigation has been performed by Bulou *et al.*[19] However, the perturbing effects of e-ph coupling (Eq.6) have never been taken into account. Aim of this Section is to determine a reference force field $\Phi^o$, in the absence of linear e-ph coupling. Therefore $\Phi^o$ is not chosen to fit the $\omega_s$ and $\omega_3$ frequencies and the problem is highly undetermined. To overcome this problem we adopt a molecular spectroscopist approach, following the Wilson's internal coordinates (**GF**) scheme.[20]

We start by constructing a force field for the isolated complexes, such as $[Pt(en)_2]Cl_2$ or $[Pt(NH_3)_4Cl_2]Cl_2$, for which extensive vibrational data are available in the literature.[21] We are not interested in the vibrations inside the ligand groups, therefore we simplify the structure of the complexes as shown in Fig. 3, in order to get a sensible description of the Pt-N vibrations, the only ones which appreciably couple to Pt-Cl motions. In the figure $N^*$ and $C^*$ are fictitious atoms, with the mass of $NH_2$ and $CH_3$, respectively: In such a way we account for the effect of the H atoms on the N and C motions, without considering the motions of the H themselves. We omit the details of the refinement process that, on the basis of the vibrational data available for the isolated complexes, yielded to the force field reported in Table II. We just mention that we made reference to the frequencies of both $Pt^{II}$ and $Pt^{IV}$ complexes, assuming that the ligand's force constants do not depend on the oxidation state of the metal.

Starting from the isolated complexes, four additional internal coordinates are required to describe the $q=0$ vibrations of the linear MX chain. Two of them correspond to the motion of Cl atoms out of the line joining the Pt atoms, and a third to the torsion of two neighbor $Pt(N^*C^*)_4$ units. Due to their high masses, vibrations involving these coordinates are probably located below 100 $cm^{-1}$, and coupled to the lattice (interchain) modes. Since there is no reliable experimental data to compare with, we arbitrarily attribute to the diagonal force constants corresponding to these three internal coordinates a value of 0.1 mdyn A/rad$^2$, and neglect all off-diagonal interactions. The fourth internal coordinate of the linear chain is the $Pt^{II} - Cl$ stretching coordinate. Of course, we need to fix all the force constants involving this new coordinate. In particular, by disregarding stretching/bending, bending/bending, as well as stretching/stretching interactions beyond the first nearest neighbor, we need the following force constants: the diagonal $Pt^{II} - Cl$ stretching ($K_4$), and the two nearest-neighbor interactions: $Pt^{IV} - Cl/Pt^{II} - Cl(F_6)$, and $Pt^{II} - Cl/Pt^{II} - Cl(F_7)$. These three force constants, together with $K_3(Pt^{IV} - Cl$ stretching) and $F_5(Pt^{IV} - Cl/Pt^{IV} - Cl$ stretching interaction) determine $\omega_a, \omega_s^o$, and $\omega_3^o$, but only the first frequency, being decoupled from the electron system, is experimentally accessible. We have then adopted the following approximations. Since the $Pt^{IV} - Cl$ distance does not change appreciably in the three compounds (as well as in the other known PtCl chains), $K_3$ and $F_5$ are set to the same value as for the isolated complex. We expect that $F_5 > F_6 > F_7$, and since appreciably in the series of the three compounds considered here (we set it to 0.12 mdyn/Å), the value of $K_4(Pt^{II} - Cl$ stretching) is completely determined by the experimental $\omega_a$ value. Tables II and III collect the values of the force constants relevant to the reference force field for complexes (1)-(3). Table IV reports the reference (calculated) and experimental frequencies for the longitudinal vibrations of the PtCl chains.

The reference force field we have constructed is also a guide in the assignment of other low-frequency modes of the studied compounds. This task is of little value for the Raman spectra, which show only few bands. The complete assignment of the IR spectra in the $100 - 500 \; cm^{-1}$ region is reported in Table V. Unfortunately, we are not able to identify the IR band associated with the third longitudinal mode of the PtCl chains, $\omega_3$. The assignment of this mode has already been matter of discussion: on the basis of polarized data on complex (2), Degiorgi et al.[18] associate it either with a $120 \; cm^{-1}$ or with a $137 \; cm^{-1}$ band. On the other hand, this interpretation is rejected by S.P.Love et al.[22], which propose the assignment to a weak $167 \; cm^{-1}$ band. According to our calculations, the reference frequency is around $100 \; cm^{-1}$ (Table IV), but the experimental frequency can be well below this value due to the effect of on-bond e-ph coupling. Another complication might arise from the coupling to the lattice modes or to the skeletal modes of the ligands. In our opinion a safe assignment of the $\omega_3$ mode requires the analysis of polarized spectra for a series of complexes with different ligands but identical crystallografic structure.

Whereas the proposed force field is by no means unique, it provides the required reference for the analysis of the e-ph coupling. Furthermore, it offers consistent guidelines for adaptation to different compounds. For instance, the reciprocal $Pt^{II} - Cl$ and $Pt^{IV} - Cl$ stretching force constants all fall on a straight line when plotted against the Pt-Cl distance (Fig. 4). Such behavior is precisely what it is expected and found for the stretching force constants of well established empirical force fields.[23]

## V. ESTIMATE OF THE MICROSCOPIC PARAMETERS

Going back to the experimental frequencies of the $Pt^{IV} - Cl$ stretchings, we again underline that, whereas the frequency of the antisymmetric vibration ($\omega_a$) increases with "chemical pressure" (i.e. increases from (1) to (3), see Fig.1 and Tab. IV), the frequency of the corresponding symmetric stretch ($\omega_s$) does decrease. This counter-intuitive behavior can be rationalized *only if* the effects of e-ph coupling described by Eq. 6 are accounted for (notice that the reference frequency of the symmetric stretch does increase along the series). In the following we describe a phenomenological model for the electronic structure of Pt-Cl chains which allows us to account for $\chi_v$ (i.e. for the vibrational frequencies) as well as for all the other optical data. Thus we can derive reliable microscopic parameters for the three investigated compounds.

We have already remarked[24] that the presence of evident Frank-Condon effects (long progressions in the resonance Raman spectra, marked difference between the absorption and luminescence maxima[2-4]) in the PtCl spectra indicates that the CT exciton is strongly localized. So we consider the smallest unit able to support CT excitons and retaining the complete symmetry of the chain, namely a trimeric unit $Pt^{II} \cdots ClPt^{IV}Cl \cdots Pt^{II}$. Within the trimer model the electronic problem is defined by the CT integral and by the energy difference between the states: $\phi = | \; Pt^{IV}Pt^{II}Pt^{IV} >; \psi_{\pm} = (2)^{-1/2} \{ \; | \; Pt^{III}Pt^{III}Pt^{IV} > \pm \; | \; Pt^{IV}Pt^{III}Pt^{III} > \}$. The CT integral mixes $\phi$ and $\psi_+$ states, whereas $\psi_-$ is unaffected. The resulting

$$\Psi_1 = c_1\phi + c_2\psi_+$$

$$\Psi_2 = c_2\phi - c_1\psi_+ \quad (8)$$

$$\Psi_3 = \psi_-$$

The CDW amplitude is given by $\sigma = (c_1)^2$, and the two coefficients are related by the normalization condition: $(c_1)^2 + (c_2)^2 = 1$. The solution to the electronic problem is thus described in terms of two parameters, $t$ and the CDW amplitude $\sigma$. For instance, the electronic susceptibility $\chi_v$ is:

$$\chi_v = \frac{2[(1-\sigma)\sigma]^{3/2}}{t} \quad (9)$$

and the frequency of the $\Psi_3 \longleftarrow \Psi_1$ transition, $\omega_{CT}^o$, is given by:

$$\omega_{CT}^o = 2t\sqrt{(1-\sigma)\sigma} \quad (10)$$

We then include the $Q_s$ coordinate in the usual Herzberg-Teller scheme. Since $\Psi_3$ is not coupled to other electronic states, the corresponding phonons are not perturbed by e-ph coupling. On the other hand, for the ground state ($\Psi_1$), one has to introduce displaced phonons (the $Q_s$ equilibrium position for the ground state is proportional to $1-\sigma$) with a softened frequency (Eq. 6). In this approximation we calculate the Frank- Condon factors for the CT transition and therefore evaluate the total oscillator strength of the transition as well as the maxima of the corresponding absorption and emission spectra. Moreover, the standard Albrecht theory of the resonance Raman effect[25] is adopted to calculate the relative intensities of the Raman progression bands.

The $e - sph$ coupling constant, $g_s$, can be estimated from Eq. 5 (the small dependence on the chain distortion is disregarded), whereas Section IV discusses the method followed to evaluate $\omega_s^o$. Therefore the parameters entering the $e - sph$ hamiltonian (Eq. 4) are known, and the two electronic parameters, $t$ and $\sigma$, can be evaluated through a best fit of the model to the available spectroscopic data. The results of the fit, together with the corresponding microscopic parameters, are reported in Tab. VI.

## VI. DISCUSSION AND CONCLUSIONS

From Table VI it turns out that the phenomenological trimer model satisfactorily accounts for the available optical data of the three compounds at hand. As a consequence we get reliable estimates of the corresponding microscopic parameters, and in particular of the CDW amplitude. The latter indeed correlates well with the Pt-Pt distance, that recent LDA calculations[26] indicate as the key parameter controlling the CDW in MX chains. Fig. 5 shows that a linear correlation exists between the $\sigma$ values from Table III (stars) and the reciprocal of the Pt-Pt distance. Since the $Pt^{IV} - Cl$ distance is practically constant (Table I), a linear relationship ($\sigma = 1.834 - 1.286r$) also exists between $\sigma$ and the ratio $r$ between the Pt-Cl distances often quoted in the literature[4]

The estimate of the CDW amplitude is of paramount importance to determine the properties of MX chains. It is therefore interesting to investigate if the diagram in Fig. 5 applies to other members of the from the available structural and vibrational data. In particular, the diagram in Fig. 4, that shows the linear relationship between $1/K_{\text{PtCl}}$ vs. $d_{\text{PtCl}}$, allows us to evaluate $K_4$ for the PtCl complexes listed in Table I ($K_3$ is constant) and therefore the reference frequencies $\omega_s^o$. We then evaluate $\sigma$ by interpolating the differences between the reference frequencies and the experimental frequencies[2-4] ($\omega_s$), and the $\sigma$ values estimated for the compounds (1) - (3) (data in Tab. IV and VI). The results are shown as squares in Fig. 5: with the exception of one compound ($[Pt(\text{etn})_4][Pt(\text{etn})_4Cl_2]Cl_4 4H_2O$, whose structural data are less accurate[4]), the points fall on the straight line determined above. Analogous results are obtained if the difference between the two experimental frequencies, $\omega_s$ and $\omega_a$, is used to estimate $\sigma$. Thus the thorough analysis of the present paper suggests simple but reliable and rather accurate methods to estimate the CDW amplitude from a limited set of experimental data, either structural or vibrational or both.

The understanding of the role of e-ph coupling allows one to explain not only the zone-center spectroscopic data of MX solids, but also some features of the phonon dispersion curves. Through accurate and detailed Raman and IR spectra of isotopically enriched $[Pt(en)_2][Pt(en)_2Cl_2](ClO_4)_4$, Love et al.[22,27] have unambiguously demonstrated that the complex structure of the Raman active Pt-Cl stretching mode $\omega_s$ (not evidenced in the present study due to the low resolution employed) is due to a Cl isotope effect. Analysis of the data gives information on the width of the phonon dispersion curves: the $\omega_s$ should disperse upward by $6.5 cm^{-1}$, from zone center to boundary, whereas the dispersion of the corresponding IR active mode, $\omega_a$, should be less than 3 $cm^{-1}$.[27] No explanation is offered for this different behavior, and to reproduce it Love et al.[22,27] propose a force field with unphysical values for the stretch-stretch interactions ($F_5 - F_7$ in Table III). On the other hand, we know that the $\omega_s$ mode is coupled to the electron system, whereas $\omega_a$ is not. We believe that the reference (unperturbed) frequency $\omega_s^o$ has a similar dispersion as the $\omega_a$ mode. On the other hand, the actual dispersion of the Raman frequency is governed by e- ph coupling: $\omega_s(q)$ is given by a generalization of Eq. 6, with the zone- center $\chi_v$ replaced by the $\chi_v(q)$ function. The $q$-dependence of the susceptibility $\chi_v$ cannot be calculated by the trimer model adopted before, but its qualitative behavior is easily understood on simple physical grounds. It is well known that $\chi_v(q)$ of the regular, undistorted chain has a divergence at $q = 2k_F$, as it results from the conventional treatment of the Peierls instability yielding the CDW state.[28] For a half-filled chain $2k_F$ corresponds to the zone boundary, which becomes the zone center after the distortion has taken place. Therefore $\chi_v(q)$ is expected to be maximum at the zone center, as a reminiscence of the Peierls instability, and correspondingly $\omega_s(q)$ is minimum and increases towards the zone edge, as experimentally found.[27]

In the present paper we have shown that the optical properties of MX chains, and particularly the opposite behavior of IR and Raman frequencies under physical or chemical pressure, can only be understood if e-ph coupling is properly accounted for. We have adopted a simple phenomenological model, but any realistic model for the electronic structure of MX chains has to include $e - sph$ coupling. In particular, Alouani et al.[26] have recently fitted several properties of PtX chains under uniaxial pressure by neglecting $e - sph$ coupling and including a hard- core Pt-X repulsion. We believe that, whereas this repulsion is important to account for the nearly constant Pt-X distance, a fit including both Raman and IR $Pt - X$ stretching frequencies would impose sizable $e - sph$ coupling strength.

compounds in the PtCl series, given their high $\sigma$ values. When intermediate $\sigma$ values are involved, as in the PtBr and PtI series,[2-4] the trimer model is expected to fail, and should be replaced by more sophisticated models such as the one- or two-bands Hubbard models. However, the treatment of the e-ph coupling we have described in the present paper remains applicable, as well as the methods followed to describe the phonon part and to construct the reference force field.

ACKNOWLEDGMENTS

Financial support by the Italian Ministry of the University and Scientific and Technological Research (M.U.R.S.T.) and by the National Research Council (C.N.R.) is acknowledged. We thank B.I. Swanson and B.Scott for communicating the results of their work before publication and for useful discussions.